\documentclass[prl,twocolumn,aps,superscriptaddress,showpacs,floatfix]{revtex4} 
\usepackage{amsmath}
\usepackage{amsfonts}
\usepackage{epsfig}

\begin{document}
\title{Fluctuations of $1/f$ noise and the low frequency cutoff paradox} 
\author{Markus \surname{Niemann}}
\email{markus.niemann@uni-oldenburg.de}
\affiliation{Institut f{\"{u}}r Physik, Carl von Ossietzky Universit{\"{a}}t Oldenburg, 26111 Oldenburg, Germany}
\author{Holger \surname{Kantz}}
\affiliation{Max-Planck-Institut f\"{u}r Physik komplexer Systeme, %
N\"{o}thnitzer Stra\ss{}e~38, 01187 Dresden, Germany}
\author{Eli \surname{Barkai}}
\affiliation{Department of Physics, Bar Ilan University, Ramat Gan 52900, Israel}

\begin{abstract}
Recent experiments on blinking quantum dots and 
weak turbulence in liquid crystals reveal the fundamental connection
between $1/f$ noise and power law intermittency. The non-stationarity of
the process implies that the power spectrum is random -- a manifestation
of weak ergodicity breaking. Here we obtain the universal distribution 
of the power spectrum, which can be used to identify intermittency 
as the source of the noise. We solve an outstanding  paradox on the  
non integrability of $1/f$ noise and the violation of Parseval's theorem.
We explain why there is no physical low frequency cutoff and 
therefore cannot be found in experiments.
\end{abstract}


\maketitle

The power spectrum $S(f)$  of a wide variety of physical systems exhibits 
enigmatic $1/f$ noise \cite{Keshner82,Hooge72} at low frequencies,  
\begin{equation}
S(f) \sim \frac{\mbox{constant}}{f^\gamma} \text{ where } 0<\gamma<2.
\label{eq01}
\end{equation}
Starting with Bernamont \cite{Bernamont37}, in the context of resistance fluctuations
of thin films, many models of these widely observed phenomena were put forward.
Indeed, $1/f$ noise is practically universal, ranging
from voltages and currents in vacuum tubes, diodes and transistors, 
to annual amounts of rainfall, to name only a few examples. 
A closer look at the phenomenon 
reveals several themes which demand further explanation.
The first is that $1/f$ noise is not integrable: $\int_{-\infty} ^\infty S(f) \, {\rm d} f=\infty$,
due to the low frequency behavior,
 when $\gamma \ge 1$.  
This violates the Parseval theorem from which one may deduce that the spectrum
 of a random process is integrable (see details below). 
So how can we find
in laboratory $1/f$ noise if a mathematical theorem forbids it? One simple
explanation would be that the phenomenon has a cutoff at some low frequency,
namely that below $f<f_0$ Eq. (\ref{eq01}) is not valid. Experimentalists
have therefore carefully searched for this cutoff, increasing the measurement time
as far as reasonable:  three weeks for noise in MOSFET \cite{Caloyannides74}, and 
$300$ years for weather data \cite{MandelbrotWallis69}. 
No cutoff frequency is observed even after these long measurement times. 
This is one of the outstanding features of $1/f$ noise.
A second old controversy, related to the first, is the
suggestion of Mandelbrot \cite{Mandelbrot67} that models of $1/f$ noise
for $\gamma\ge 1$ should be related to non-stationarity processes, though the nature
of this non-stationarity is still an open question. 
Further, experiments find that at least in some cases the
amplitude of the power spectrum varies among identical systems measured
at different times, but the shape and in particular the
value of the exponent $\gamma$ is quite consistent \cite{Hooge72,Keshner82}. 
This means that a $1/f$ spectrum is a non self averaging observable,
at least in some systems. 

While these observations where made long ago, the verdict on them is not yet
out.  
However, recent measurements of blinking
quantum dots \cite{PeltonGrierGuyot-Sionnest04,PeltonSmithSchererEtAl07} 
and liquid crystals in the electrohydrodynamic convection regime \cite{SilvestriFronzoniGrigoliniEtAl09}
shed new light on the nature of $1/f$ noise.
Both systems, while very different in their nature, reveal a 
power law intermittency route to $1/f$ noise.
This means that power law waiting times in a micro-state of the system
are responsible for the observed spectrum.
This approach was suggested as a fundamental mechanism
for $1/f$ noise in the context of intermittency of chaos and
turbulence, with the work of Manneville \cite{Manneville80}. 
Subsequently, it has been found in many intermittent chaotic systems 
\cite{GeiselZacherlRadons87, ZumofenKlafter93a, Ben-MizrachiProcacciaRosenbergEtAl85, GeiselNierwetbergZacherl85},
and has been used successfully as a model 
for transport in geological formations \cite{ScherMargolinMetzlerEtAl02}.
For a quantum dot driven by a continuous wave laser,
this mechanism means that the dot switches from
a dark state to a bright state where photons are emitted, and that
sojourn times in both states exhibit power law statistics which is scale
free \cite{Fransuzov08, StefaniHoogenboomBarkai09}. 
Waiting times probability density functions (PDF)
in these states follow  $\psi(\tau) \sim \tau^{-(1+\alpha)}$ and $0<\alpha<1$
(for bright state times this behavior is found for low laser intensity and low temperature). 
The dynamics is scale free, because the average sojourn times
diverge, and we expect weak ergodicity breaking \cite{Bouchaud92,BrokmannHermierMessinEtAl03}.
This means that the  power spectrum remains a random variable even in the
long time limit \cite{MargolinBarkai05,MargolinBarkai06}. 

Here we investigate the non self averaging power spectrum, and
show that indeed this observable exhibits large but universal fluctuations,
while the estimation of $\gamma$ is rather robust. 
Our work gives experimentalists
a way to verify whether a data set exhibiting $1/f$ noise belongs
to the intermittency class, and this we believe will  help unravel the origin
of an old mystery of statistical physics.  
We also remove the paradox based on Parseval's identity, 
showing that as $t \to \infty$ the integrability remains, and
that there is no cutoff frequency $f_0$. So experimentally
searching for this ``lost" low frequency might be in vain. 

{\em Parseval's identity and $1/f$ noise}. We consider a measurement of
a random signal $I(t)$ in the time interval $(0,t)$, so that its Fourier transform
is $\tilde{I}_{t} (\omega) = \int_0 ^t I(t') \exp(- i \omega t'){\rm d} t'$.
The power spectrum
$S_{t} \left( \omega \right) = [ \tilde{I}_t \left( \omega \right) \tilde{I}^{*}_t \left( \omega \right) ] / t$,
is considered in the long measurement time limit. The ensemble average power spectrum
is $\langle S_t \left( \omega \right) \rangle$.
Note that in an experiment with one realization of the time
series, for example a measurement of the intensity of 
a single molecule or a quantum dot, the ensemble average
is not performed, though in experiments one introduces smoothing 
methods which reduce the noise level of the 
reported power spectrum \cite{numericalRecipes}.
More importantly note that the integral over the power spectrum is
\begin{widetext}
\begin{equation}
 \int_{-\infty} ^\infty S_t \left(\omega \right) {\rm d} \omega =  
\frac{1}{t} \int_{-\infty} ^\infty {\rm d} \omega
\int_0 ^t {\rm d} t_1 \, \exp\left(- i \omega t_1\right) I(t_1) 
\int_0 ^t {\rm d} t_2\, \exp\left(i \omega t_2\right) I(t_2)
=
\frac{2 \pi}{t} \int_0 ^t I^{2} (t_1) {\rm d} t_1,
\label{eq02}
\end{equation}
\end{widetext}
where we used a well known identity of the delta function 
$\int_{-\infty} ^\infty {\rm d} \omega \exp[-i \omega(t_1-t_2)] = 2 \pi \delta(t_2 - t_1)$ and $S_t(w) = S_t (-\omega)$ by definition. 
For any bounded process  be it ergodic or non-ergodic, stationary or non-stationary,
$I^{2}(t) \le (I_\mathrm{max})^2$ and hence 
$\int_{-\infty} ^\infty S_t \left(\omega \right) {\rm d} \omega \le 2 \pi (I_{\max})^2$. So
the integral is finite for a wide class of processes. As a consequence,
non-integrable $1/f$ noise is strictly prohibited. 
The classical way out was to {\sl assume} 
a violation of the $1/f$-behaviour in the limit of $f\to 0$. 

The ensemble-average of Eq. \eqref{eq02}, under the additional
assumption that the process reaches a stationary state, reads 
$\lim_{t \to \infty} \int_{-\infty} ^\infty S_t (\omega) {\rm d} \omega = 2
\pi \langle I^2 \rangle$. If the system is ergodic, i.e., $\overline{I^2} = \int_0 ^t I^2(t') {\rm d} t'/t \to \langle I^2 \rangle$,
we have for a single trajectory $I(t')$
\begin{equation}
\lim_{t \to \infty} \int_{-\infty} ^\infty S_t \left(\omega \right) {\rm d} \omega = 2 \pi \langle I^2 \rangle.
\label{eq03}
\end{equation}
Thus fluctuations of the total area under the power spectrum are an indication
for ergodicity breaking.

{\em Models.}  For simplicity we consider a two state model. The
generalization to $N$ states will be discussed elsewhere. 
The two states of our model are \textit{up} or $+$
 where $I(t)=I_0$ and \textit{down} or $-$
with $I(t)=-I_0$. The sojourn times in these states are independently 
identically distributed
random variables with PDFs $\psi (\tau)$. Thus after waiting a random time in
state \textit{up}, the particle jumps to state \textit{down} and vice versa. The waiting times PDFs
have long tails $\psi(\tau) \propto \tau^{-(1 + \alpha)}$ with $0<\alpha<1$,
hence the averages of \textit{up} and \textit{down} times are infinite. 
The Laplace $t \to \lambda$ transform
of these PDFs is for small $\lambda$: $\hat{\psi}(\lambda) \simeq 1 - (\overline{\tau} \lambda)^\alpha$
where $\overline{\tau}$ is a scaling constant. This is a simple stochastic model of a blinking quantum dot,
for which typically $\alpha = 1/2$ though $1/2<\alpha<1$ was also reported.  

More general models where the random process $I(t)$ has $N$ internal states
with possible different waiting time distributions
can describe annealed trap models used for glass phenomenology
\cite{Bouchaud92} or for continuous time random walks describing motion of
single molecules in live cells \cite{BarkaiGariniMetzler12}. Our results 
for $N>2$ are structurally similar to those for $N=2$ but deserve their own
discussion, which will be presented in a longer paper. For these models,
we have derived detailed expressions for the power spectrum 
of the process $I(t)$ and its statistical properties.


{\em Statement of the main results.} For $\alpha < 1$ the expectation
value of the spectrum is not constant, but decreases with measurement time
$\langle S_t(\omega) \rangle \simeq t^{\alpha - 1} \sigma_\alpha(\omega)$.
Expanding the $t$-independent function $\sigma_\alpha(\omega)$ for small
frequencies $\omega$, one finds a typical non-integrable $1/f$-noise
\begin{equation}
\langle S_t(\omega) \rangle \simeq C \frac{t^{\alpha - 1}}{\omega^{2-\alpha}}.
\label{eq:expectedSpectrum}
\end{equation}
In general, the value $S_t(\omega)$ of the
spectrum is a fluctuating quantity even in the $t \to \infty$ limit.
The statistical behavior of the general class of processes 
for large $t$ (for pairwise disjoint $\omega_i \neq 0$) is fully described by 
\begin{equation}
\left(\frac{S_t (\omega_1)}{\langle S_t (\omega_1) \rangle}, 
      \dotsc, \frac{S_t (\omega_n)}{\langle S_t (\omega_n) \rangle} \right) 
\to Y_\alpha \cdot (\xi_1, \dotsc, \xi_n),
\label{fractionalPeriodogram}
\end{equation}
where $Y_\alpha$ is a random variable of normalized Mittag-Leffler distribution with 
exponent $\alpha$ whose moments are
$\langle Y_\alpha^n \rangle = n! \Gamma(1+\alpha)^n / \Gamma(1+n\alpha)$ \cite{Feller71}.
The $\xi_i$ are independent exponential random variables with unit mean.
For $\alpha = 1$ the Mittag-Leffler random variable becomes $Y_1 = 1$, so that
the powers $S_t(\omega_i)$ of 
different frequencies become independent exponentially distributed
random variables - 
a result known for several ergodic random processes \cite{Priestley81}.
In the case of weak ergodicity breaking ($\alpha < 1$), the whole spectrum
obtains a common random prefactor $Y_\alpha$ which shifts the complete 
observed spectrum.

Many procedures for the estimation of the spectrum from one finite time 
realization are 
designed to suppress the statistical fluctuations due to the uncorrelated 
random variables $\xi_i$ \cite{Priestley81, numericalRecipes}. These cannot
account for the fluctuations of $Y_\alpha$, common to all estimators of a 
given realization. 
For these procedures the prefactor affects all estimated values for the
spectrum. However, being a common prefactor, it does not affect the 
shape of the estimated spectrum so that features
as $1/f$-noise can be detected independently of the realization.

{\em Motivation of the results.} Whereas an exact derivation for the general $N$-state model is based on \cite{NiemannSzendroKantz10}
and will be presented in a longer paper, we report here on the 
2-state model introduced above. To simplify our arguments further, we 
assume that after a
waiting time the next state will be chosen randomly (i.e., a $+$ state with probability $1/2$).

Let $\tau_i$ be the $i$th waiting time and $\chi_i= \pm I_0$ the value
taken during this waiting time. We denote by 
$T_j = \sum_{i=1}^{j - 1} \tau_i$ the times at which one waiting
time ends. If $n(t)$ is the number of completed waiting times up to time $t$,
we can approximate by ignoring the waiting time in progress at $t$
\begin{multline}
\int_0^t \mathrm{d}\tau \, \exp(i\omega \tau) I(\tau) 
\simeq \sum_{j=1}^{n(t)} d_j(\omega) \\
\text{with } d_j(\omega) = i \chi_j \exp(i \omega T_j) \frac{1 - \exp(i \omega \tau_j)}{\omega}.
\end{multline}
With this approximation and $\langle \chi_i \chi_j \rangle = \delta_{ij} I_0^2$ one
obtains 
\begin{equation}
S_t(\omega) \simeq \frac{1}{t} \sum_{k,l=1}^{n(t)} d_k(\omega) d_l(-\omega).
\label{eq:approxSpectrum}
\end{equation}
Assuming that $n(t)$ is for large $t$ independent of the
waiting time of a single step, $\tau_i$, we get for the ensemble average
\begin{equation}
\begin{split}
\langle S_t(\omega) \rangle 
&\simeq \frac{\langle n(t) \rangle}{t} \langle d_1 (\omega) d_1(-\omega) \rangle \\
&\simeq I_0^2 \frac{\langle n(t) \rangle}{t} \frac{2 - \hat{\psi}(i\omega) - \hat{\psi}(-i\omega)}{\omega ^2}.
\end{split}
\label{eq:firstEnAv}
\end{equation}
It has been shown that 
$n(t) \simeq Y_\alpha t^\alpha / (\Gamma(1+\alpha) \overline{\tau}^\alpha )$ 
\cite{BouchaudGeorges90}. Therefore
\begin{equation}
\begin{split}
\langle S_t(\omega) \rangle
&\simeq \frac{I_0^2 t^{\alpha-1}}{\Gamma(1+\alpha) \overline{\tau}^\alpha}
        \frac{2 - \hat{\psi}(i\omega) - \hat{\psi}(-i\omega)}{\omega ^2} \\
&\simeq \frac{2 I_0^2 \cos(\alpha\pi/2)}{\Gamma(1+\alpha)} \frac{t^{\alpha-1}}{|\omega|^{2-\alpha}} \quad \text{as } \omega \to 0.
\end{split}
\end{equation}
The last line shows the typical $1/f$ noise \cite{MargolinBarkai06}.
It is important that the observation limit $t \to \infty$ is taken before the
frequency limit $\omega \to 0$.

We motivate the main result Eq.~\eqref{fractionalPeriodogram} with help
of a random phase approximation.
The random phase approximation assumes that terms of the form $\exp(i \omega T_j)$ are just
random phases and any average over them vanishes. Especially, 
$\langle d_{j_1}(\nu_1) \dotsm d_{j_n}(\nu_n) \rangle = 0$ if 
$\nu_1 T_{j_1} + \dotsb + \nu_n T_{j_n} \neq 0$ for some $T_j$s (the $\nu$ being
$\pm \omega$). 
Looking at the second moment of Eq.~\eqref{eq:approxSpectrum} and using \eqref{eq:firstEnAv}:
\begin{equation}
\begin{split}
\left\langle S_t^2(\omega) \right\rangle  
&\simeq \frac{1}{t^2} \langle \sum_{k,l,p,q=1}^{n(t)} d_k(\omega) d_l(-\omega) d_p(\omega) d_q(-\omega) \rangle \\
&\simeq \frac{2}{t^2} \langle n(t)^2 \rangle \langle d_1(\omega) d_1(-\omega) \rangle^2 \\
&\simeq 2 \langle Y_\alpha^2 \rangle \langle S_t(\omega) \rangle^2
\end{split}
\label{derivationMoment}
\end{equation}
where we ignored terms with $k=l=p=q$ as there are only $\langle n(t) \rangle$ of them. The
factor $2$ stems from the fact that the sum in the first line of Eq.~\eqref{derivationMoment}
has contributions for $k=l$, $p=q$ and for $k=q$, $l=p$.
In contrast to this, for the term 
$\langle S_t(\omega_1) S_t(\omega_2) \rangle$ with $\omega_1 \neq \omega_2$ this
symmetry factor will not be present. Following the same steps as in Eq.~\eqref{derivationMoment}
gives 
\begin{equation}
\langle S_t(\omega_1) S_t(\omega_2) \rangle
\simeq \langle Y_\alpha^2 \rangle \langle S_t(\omega_1) \rangle \langle S_t(\omega_2) \rangle.
\end{equation}
This shows the equality of the second moments of Eq.~\eqref{fractionalPeriodogram}.
The equality of the higher moments follows similarly 
by using combinatorial methods to determine these symmetry factors. 
We see that the random number of jumps $n(t)$ is responsible for the Mittag-Leffler fluctuations
while the random phases generate the exponential noise. 
See Supplemental Material at [URL will be inserted by publisher] for a more detailed version
of this approximation.
The exact proofs for the general model will be published in a longer paper.

\begin{figure}
\centerline{\includegraphics[width=\columnwidth]{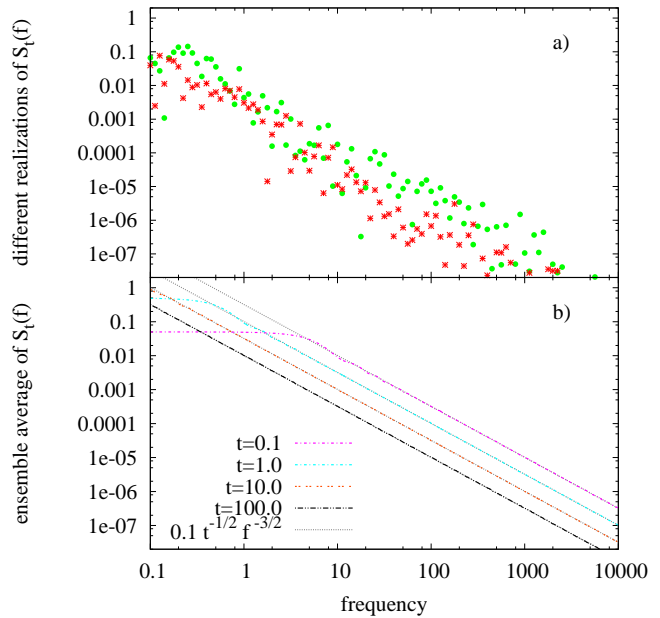}}
\caption{(color online) a) $S_t(f)$ plotted for different realizations ($\alpha=0.5$, $t=10^2$).
Inside each realization one has fluctuations following exponential distributions.
Different realizations are shifted with respect to each other due the random prefactor $Y_\alpha$.
b) Ensemble average of $S_t(f)$ plotted for different lengths $t$ of the time series.
One sees the decay of the spectrum $\langle S_t(f) \rangle \simeq 0.101\, t^{-1/2} f^{-3/2}$ both in time and frequency.
The crossover frequency is around $f_\mathrm{c} \simeq 0.51/t$. The simulation perfectly
match the theory Eqs.~(\ref{eq:expectedSpectrum},\ref{eqbbb}).
\label{averageSpectrum}}
\end{figure}

{\em Numerical results.} We simulated the two state model with $I_0=1$ for different
length of time series and different $\alpha$. The waiting times were generated
by using a uniformly distributed random number $0 < X \leq 1$ and setting $\tau = c_\alpha X^{-1/\alpha}$.
The constant $c_\alpha$ was chosen such that $\langle n(1) \rangle \simeq 10\,000$.
The ensemble consists of $10\,000$ realizations of the time series.

In Fig.~\ref{averageSpectrum}a we have plotted different realizations of $S_t(f)$.
The stochastic fluctuations inside and between the realizations are
clearly observable. In Fig.~\ref{averageSpectrum}b the ensemble average of the power
spectrum for different lengths is plotted. The $1/f$ spectrum and its decay with
observation time is clearly visible. Note that at very low frequencies we find 
$S_t(\omega) \simeq \mathrm{const}$ independent of frequency - an effect we soon explain.

\begin{figure}
\centerline{\includegraphics[width=\columnwidth]{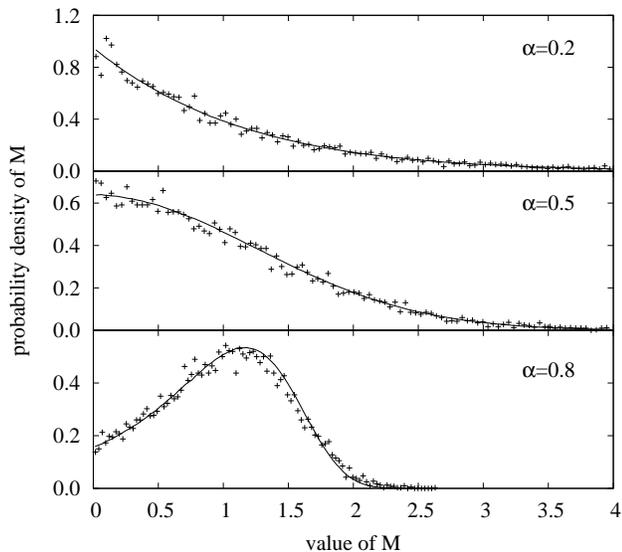}}
\caption{Distributions of the frequency averaged spectra M (see Eq.~\eqref{frequencyAverage}).
The lines are the analytic probability densities of the Mittag-Leffler distributions
($t=10^4$)
\label{mldistributions}}
\end{figure}

In a second step we want to check the statistical properties described
by Eq.~\eqref{fractionalPeriodogram}. To isolate the Mittag-Leffler
fluctuations, we have calculated the spectrum for a fixed set of
$N$ frequencies $\omega_i$ and determined the values
\begin{equation}
M = \frac{1}{N} \sum_{i=1}^N \frac{S_t(\omega_i)}{\langle S_t(\omega_i) \rangle}.
\label{frequencyAverage}
\end{equation}
As the exponential distributions are uncorrelated, they average out for
sufficiently large $N$ and the value
taken by $M$ should be distributed as the Mittag-Leffler distribution
$Y_\alpha$. We have compared this for different $\alpha$ values. The
histogram of $M$ values with the Mittag-Leffler density is shown
in Fig.~\ref{mldistributions} for $\alpha = 0.2$, $\alpha = 0.5$
and $\alpha = 0.8$. A good agreement with the theory is apparent. 


{\em Removing the non integrability paradox of $1/f$ noise.}
As mentioned in the introduction, the $1/\omega^{2-\alpha}$
noise is non integrable, 
$\int_0 ^\infty \langle S(\omega) \rangle {\rm d} \omega =\infty$,
due to the low frequency behavior. This in turn violates the simple
bound we have found. To start understanding this behavior
notice that the random phase approximation breaks down when $\omega=0$,
as the phase $\omega T_n$ is clearly non random. 
Hence the distribution of the power spectrum in Eq.~\eqref{fractionalPeriodogram}
is not valid for $\omega=0$ and this case must be treated separately.

For $\omega=0$ we have for a single realization
$S_t(0) = \overline{I}^2 t$ with the time average 
$\overline{I} = \int_0^t I(t') {\rm d} t'/t$. 
For ergodic processes the time average $\overline{I}$ is equal
to the ensemble average $\langle I \rangle$. However, for non ergodic
processes under investigation, the time average $\overline{I}$ remains
a random variable even in the infinite time limit \cite{MargolinBarkai05, RebenshtokBarkai07}. 
For the two state process introduced above,
we have $\overline{I} = I_0 ( T^{+} - T^{-}) / t $ where 
$T^{\pm}$ is the total time spent in state \textit{up} or \textit{down}.
The value of $\overline{I}$ follows an
arcsine like distribution \cite{GodrecheLuck01,MargolinBarkai05,RebenshtokBarkai07,RebenshtokBarkai08}. 
This simply means that for a given realization, the system will spend 
most of the time either in state \textit{up}, or in state \textit{down}, and
hence $\overline{I}$ is random, which would not be the case for an ergodic process.

This has a consequence for the non integrability of the
power spectrum. As $S_t(0) = \overline{I}^2 t$,
the spectrum at zero frequency tends to infinity, but for any
finite measurement time it is finite. For the two state model
we have on average $\langle S_t(0) \rangle = I_0^2 (1 - \alpha) t$. 
So indeed \emph{theoretically} there is 
a low-frequency cut-off of the divergence of the $1/f$ spectrum, and we
 now define a crossover frequency $\omega_{\mathrm{c}}$ for 
 the transition between the zero frequency limit,
where arcsine statistics takes control (failure of random phase
approximation), and higher frequencies where the Mittag-Leffler statistics
takes control. 
This frequency is defined by merging the two behaviours, 
\begin{equation}
\langle S_t(\omega_\mathrm{c}) \rangle \simeq C \frac{t^{\alpha-1}}{\omega_\mathrm{c} ^{2 - \alpha}} 
= \langle S_t(0) \rangle.
\label{eqbbb}
\end{equation}
%
We see that
\begin{equation}
\omega_\mathrm{c} = \left( C/\langle \overline{I}^2 \rangle \right)^{1/(2-\alpha)}
 \frac{1}{t}.
\end{equation}
The values of $\langle \overline{I}^2 \rangle$ and $C$ can be obtained
from measurements, or from theory, for example for the two state model
$\langle \overline{I}^2 \rangle= I_0^2 (1 - \alpha)$ and $C = 2 I_0^2 \cos(\alpha \pi /2) /\Gamma(1+\alpha)$.
More importantly, we see that the crossover frequency depends on the measurement
time as $1/t$. Although at first sight surprising, this is the only way how
such a cross-over can take place in the absence
of a characteristic time scale for dynamics: measurement time itself sets
the time scale for crossover. Additionally, $1/t$ appears as the frequency
resolution of the discrete Fourier transform typically used in spectral
analysis. Importantly, experiments report a lowest frequency at $f=1/t$.

We see that increasing measurement time merely stretches the domain
of frequency where the $1/f$ noise is observed, which is clearly seen
in the numerical simulations (see Fig.~\ref{averageSpectrum}b). 
There is no point in increasing measurement time in order to identify better
the crossover, since 
a time independent crossover frequency does not exist.
Thus the $1/f$ noise stretches to the lowest frequencies compatible
with measurement time (of the order of $1/t$). This resolves the
non integrability paradox. The amplitude of the power spectrum itself is also
decreasing in time, in such a way that integrability is maintained.
Namely
\begin{equation}
\begin{split}
\int_0^\infty \langle S_t(\omega) \rangle {\rm d} \, \omega 
&\simeq \langle \overline{I}^2 \rangle t \omega_\mathrm{c}
  + \int_{\omega_c}^\infty C t^{\alpha-1} / \omega^{2-\alpha} \, {\rm d} \omega  \\
&= \frac{2-\alpha}{1-\alpha} \left( \langle \overline{I}^2 \rangle^{1-\alpha} C \right)^{1/(\alpha-2)}
\end{split}
\label{eqmmm}
\end{equation}
is indeed finite and time independent.

Thus we conclude that the power spectrum is integrable as it should. 
This seems to indicate the generality of our results, since a crossover
frequency is only found in few experiments. 
From a different angle, 
assuming that the natural frequency is also the limit of measurement $\omega_c \sim 1/t$,
we must demand the decrease of the amplitude of power spectrum with time
to maintain integrability as required for bounded signals. 
This together with the universal fluctuations of $1/f$ noise (Eq.~\eqref{fractionalPeriodogram})
are strong fingerprints of power law intermittency. The tools developed here can be
tested in a vast number of physical systems.

{\em Acknowledgments.} 
This work is supported by the Israel Science Foundation.
M.N. thanks Daniel Nickelsen for a careful reading of the manuscript.
The calculations were carried out the HERO
(High-End Computing Resource Oldenburg) at the University of Oldenburg.

\end{document}


\title{Supplementary material to: \\
Fluctuations of $1/f$ noise and the low frequency cutoff paradox} 
\author{Markus \surname{Niemann}}
\author{Holger \surname{Kantz}}
\author{Eli \surname{Barkai}}
\noaffiliation

\maketitle

\newcommand{\dop}{\mathrm{d}}
\renewcommand{\theequation}{S\arabic{equation}}

This supplementary material contains a more detailed derivation of the
main results. We will still work with the approximations made in the
main text -- an exact proof will be published in a longer paper.

Using the same notations as in the main text we see
\begin{equation}
\begin{split}
d_j(\omega)
&= \int_{T_j}^{T_{j+1}} \dop \tau \, \chi_j \exp(i\omega \tau) \\
&= i \chi_j \exp(i \omega T_j) \frac{1 - \exp(i \omega \tau_j)}{\omega}.
\end{split}
\end{equation}
With this, we get for the Fourier transform of the signal by
neglecting the last waiting time in progress at $t$:
\begin{equation}
\begin{split}
\int_0^t \mathrm{d}\tau \, \exp(i\omega \tau) I(\tau) 
=& \sum_{j=1}^{n(t)} \int_{T_j}^{T_{j+1}} \dop \tau \, \chi_j \exp(i\omega \tau)  
  + \int_{T_{n(t)+1}}^{t} \dop \tau \, \chi_j \exp(i\omega \tau) \\
\simeq& \sum_{j=1}^{n(t)} d_j(\omega),
\end{split}
\end{equation}
where $n(t)$ denotes the number of finished waiting times up to time $t$ also 
called the number of renewals in $(0,t)$.
With this approximation we get for the spectrum
\begin{equation}
S_t(\omega) \simeq \frac{1}{t} \sum_{k,l=1}^{n(t)} d_k(\omega) d_l(-\omega).
\end{equation}
We now assume that the $T_i$ are independent from the $\tau_j$ and in turn
the average of $n(t)$ can be performed independently from the average over the
$\tau_j$.
Using $\langle \chi_i \chi_j \rangle = \delta_{ij} I_0^2$ one
obtains (where a distinction is necessary, we write the variables whose average is taken as an index)
\begin{equation}
\begin{split}
\langle S_t(\omega) \rangle 
&\simeq \frac{1}{t} \langle \sum_{k,l=1}^{n(t)} d_k(\omega) d_l(-\omega) \rangle \\  
&\simeq \frac{1}{t} \left\langle \sum_{k,l=1}^{n(t)} \langle \chi_k \chi_l \rangle_{\chi_i}
                     \left\langle \exp\left( i\omega (T_k - T_l) \right) \frac{1 - \exp(i \omega \tau_k)}{\omega} \frac{1 - \exp(-i\omega \tau_l)}{\omega} 
                     \right\rangle_{\tau_i,T_i} \right\rangle_{n(t)} \\
&= I_0^2 \frac{1}{t} \left\langle \sum_{k=1}^{n(t)} \frac{\langle 2 - \exp( i \omega \tau_k) - \exp(- i \omega \tau_k) \rangle_{\tau_i}}{\omega^2} \right\rangle_{n(t)} \\
&\simeq I_0^2 \frac{\langle n(t) \rangle}{t} \frac{2 - \hat{\psi}(i\omega) - \hat{\psi}(-i\omega)}{\omega ^2} \\
&\simeq \frac{\langle n(t) \rangle}{t} \langle d_1(\omega) d_1(-\omega) \rangle.
\end{split}
\label{eq:ensembleSpectrum}
\end{equation}
It has been shown that 
$n(t) \simeq Y_\alpha t^\alpha / (\Gamma(1+\alpha) \overline{\tau}^\alpha )$ 
\cite{GodrecheLuck01,BouchaudGeorges90} from which we get 
$\langle n(t) \rangle \simeq t^\alpha / (\Gamma(1+\alpha) \overline{\tau}^\alpha )$
\begin{equation}
\langle S_t(\omega) \rangle
\simeq \frac{I_0^2 t^{\alpha-1}}{\Gamma(1+\alpha) \overline{\tau}^\alpha}
        \frac{2 - \hat{\psi}(i\omega) - \hat{\psi}(-i\omega)}{\omega ^2}.
\end{equation}
Here $Y_\alpha$ is a random variable of Mittag-Leffler distribution with unit mean
which is defined by its moments $\langle Y_\alpha^n \rangle = n! \Gamma(1+\alpha)^2 / \Gamma(1+n\alpha)$.
Using a random variable $L_\alpha$ of totally asymmetric
L\'{e}vy stable distribution with exponent $\alpha$ and distribution $l_\alpha(\xi)$
\begin{equation}
\langle \exp(-\lambda L_\alpha) \rangle = \int_0^\infty \dop \xi \, \exp(-\lambda \xi) l_\alpha(\xi) = \exp(-\lambda^\alpha),
\end{equation}
a representation of $Y_\alpha$ can be obtained by
\begin{equation}
Y_\alpha = \Gamma(1+\alpha) L_\alpha^{-\alpha}
\end{equation}
which is equivalent of $Y_\alpha$ having the probability density function \cite{BouchaudGeorges90,HeBurovMetzlerBarkai08}
\begin{equation}
y_\alpha(\xi) = \frac{\Gamma^{1/\alpha}(1+\alpha)}{\alpha \xi^{1+1/\alpha}} 
                l_\alpha\left[ \frac{\Gamma^{1/\alpha}(1+\alpha)}{\xi^{1/\alpha}} \right].
\end{equation}
Plots of $y_{\alpha}(\xi)$ are used in Fig. 2 for various $\alpha$.

For small $\lambda$ we have the expansion in the Laplace variable $\hat{\psi}(\lambda) \simeq 1 - ( \overline{\tau} \lambda )^\alpha$. 
By analytic continuation (putting the branch cut to the negative real axis such that it
is not crossed) we have
\begin{equation}
\begin{split}
(i \omega)^\alpha 
&= \left( |\omega| \exp\left[ i \frac{\pi}{2} \operatorname{sgn}(\omega) \right] \right)^\alpha \\
&= |\omega|^\alpha \left( \cos\left( \frac{\pi}{2}\alpha\right) 
   + i \operatorname{sgn}(\omega) \sin\left( \frac{\pi}{2}\alpha\right) \right)
\end{split}
\end{equation}
such that for small $\omega$
\begin{equation}
1- \hat{\psi}(i\omega) \simeq \overline{\tau}^\alpha |\omega|^\alpha \left( \cos\left( \frac{\pi}{2}\alpha\right) 
   + i \operatorname{sgn}(\omega) \sin\left( \frac{\pi}{2}\alpha\right) \right)
\end{equation}
and
\begin{equation}
\langle S_t(\omega) \rangle 
\simeq \frac{2 I_0^2 \cos(\alpha\pi/2)}{\Gamma(1+\alpha)} \frac{t^{\alpha-1}}{|\omega|^{2-\alpha}} \quad \text{as } \omega \to 0.
\end{equation}
It is important, that the observation limit $t \to \infty$ is taken before the
frequency limit $\omega \to 0$.

The Eq.~\eqref{fractionalPeriodogram} is shown by checking the equality of the
moments on both sides. Assuming that we have some natural numbers 
$p_1,\dotsc,p_q$ ($P=p_1 + \dotsb + p_q$), 
we need to show
\begin{equation}
\begin{split}
\left\langle \prod_{j=1}^q \left( \frac{S_t(\omega_j)}{\langle S_t(\omega_j) \rangle} \right)^{p_j} \right\rangle
&\overset{!}{=} \left\langle \prod_{j=1}^q (Y_\alpha \xi_j)^{p_j} \right\rangle \\
&= \langle Y_\alpha^P \rangle \prod_{j=1}^q \langle \xi_j^{p_j} \rangle \\
&= \langle Y_\alpha^P \rangle \prod_{j=1}^q p_j!.
\end{split}
\end{equation}
Here, we denote by $\overset{!}{=}$ an equality which needs to be proven.
We have used the fact that the $\xi_i$ are independent random variables which are
exponentially distributed with unit mean, i.e., $\langle \xi_i^n \rangle = n!$.
Especially, we have to show for the second moments ($\omega_1 \neq \omega_2$)
\begin{align}
\label{eq:condition1}
\left\langle \left( \frac{S_t(\omega)}{\langle S_t(\omega) \rangle} \right)^2 \right\rangle 
&\overset{!}{=} 2 \langle Y_\alpha^2 \rangle \\
\label{eq:condition2}
\text{and } 
\left\langle \frac{S_t(\omega_1)}{\langle S_t(\omega_1) \rangle} \frac{S_t(\omega_2)}{\langle S_t(\omega_2) \rangle} \right\rangle
&\overset{!}{=} \langle Y_\alpha^2 \rangle.
\end{align}

We motivate this with help of a random phase approximation.
The random phase approximation assumes that terms of the form $\exp(i \omega T_j)$ are just
random phases and any average over them vanishes. Especially, 
the phase factor of $d_{j_1}(\nu_1) \dotsm d_{j_n}(\nu_n)$ is
$\exp\left(i[ \nu_1 T_{j_1} + \dotsb + \nu_n T_{j_n} ] \right)$ 
(the $\nu_i$ take values of $\pm \omega$).
Therefore, we assume $\langle d_{j_1}(\nu_1) \dotsm d_{j_n}(\nu_n) \rangle = 0$ if 
$\nu_1 T_{j_1} + \dotsb + \nu_n T_{j_n} \neq 0$ for some $T_j$s. 
Looking at
\begin{equation}
\begin{split}
\left\langle S_t^2(\omega) \right\rangle  
&\simeq \frac{1}{t^2} \langle \sum_{k,l,p,q=1}^{n(t)} d_k(\omega) d_l(-\omega) d_p(\omega) d_q(-\omega) \rangle \\
&\simeq \frac{1}{t^2} \left\langle \sum_{k,l,p,q=1}^{n(t)} \langle d_k(\omega) d_l(-\omega) d_p(\omega) d_q(-\omega) \rangle_{\chi_i,T_i,\tau_i} \right\rangle_{n(t)} \\
&\simeq \frac{1}{t^2} \biggl\langle \sum_{k,l,p,q=1}^{n(t)}
      \langle \chi_k \chi_l \chi_p \chi_q \rangle_{\chi_i} 
      \left\langle \frac{1-\exp(i\omega\tau_k)}{\omega} \frac{1-\exp(-i\omega\tau_l)}{\omega} 
              \frac{1-\exp(i\omega\tau_p)}{\omega} \frac{1-\exp(-i\omega\tau_q)}{\omega} \right\rangle_{\tau_i} \\
&\qquad \qquad \qquad \qquad  \langle \exp( i \omega (T_k - T_l + T_p - T_q)) \rangle_{T_i} \biggr\rangle_{n(t)}.
\end{split}
\label{eq:beginSecondMoment}
\end{equation}
In our approximation, the non vanishing terms in the sum must have 
the property $\omega ( T_k - T_l + T_p - T_q) = 0$ for any realization. 
This is possible either for $k=l$ and $p=q$, or for $k=q$ and $p=l$.
In other words, the random phase approximation boils down to
\begin{equation}
\langle \exp( i \omega (T_k - T_l + T_p - T_q)) \rangle_{T_i} \simeq \delta_{kl}\delta_{pq} + \delta_{kq} \delta_{pl}.
\end{equation}
Plugging this back into Eq.~\eqref{eq:beginSecondMoment} and using \eqref{eq:ensembleSpectrum} gives
\begin{equation}
\begin{split}
\left\langle S_t^2(\omega) \right\rangle 
&\simeq \frac{2}{t^2} \left\langle \sum_{\substack{k,p=1 \\ k \neq p}}^{n(t)} \langle d_k(\omega) d_k(-\omega) \rangle \langle d_p(\omega) d_p(-\omega) \rangle \right\rangle
       + \frac{1}{t^2} \left\langle \sum_{k=1}^{n(t)} \langle d_k(\omega) d_k(-\omega) d_k(\omega) d_k(-\omega) \rangle  \right\rangle \\
&\simeq \frac{2}{t^2} \langle n(t) (n(t) - 1) \rangle \langle d_1(\omega) d_1(-\omega) \rangle^2 
       + \frac{1}{t^2} \langle n(t) \rangle \langle d_1(\omega) d_1(-\omega) d_1(\omega) d_1(-\omega) \rangle \\
&\simeq \frac{2}{t^2} \langle n(t)^2 \rangle \langle d_1(\omega) d_1(-\omega) \rangle^2 \\
&\simeq 2 \langle Y_\alpha^2 \rangle \langle S_t(\omega) \rangle^2.
\end{split}
\end{equation}
where we ignored terms of the order $\langle n(t) \rangle$ as for large $t$ the leading
order is $\langle n(t)^2 \rangle$. This shows Eq.~\eqref{eq:condition1}.

In contrast to this, for the term 
$\langle S_t(\omega_1) S_t(\omega_2) \rangle$ with $\omega_1 \neq \omega_2$ we get
\begin{equation}
\begin{split}
\left\langle S_t(\omega_1) S_t(\omega_2) \right\rangle  
&\simeq \frac{1}{t^2} \langle \sum_{k,l,p,q=1}^{n(t)} d_k(\omega_1) d_l(-\omega_1) d_p(\omega_2) d_q(-\omega_2) \rangle \\
&\simeq \frac{1}{t^2} \left\langle \sum_{k,l,p,q=1}^{n(t)} \langle d_k(\omega_1) d_l(-\omega_1) d_p(\omega_2) d_q(-\omega_2) \rangle_{\chi_i,T_i,\tau_i} \right\rangle_{n(t)} \\
&\simeq \frac{1}{t^2} \biggl\langle \sum_{k,l,p,q=1}^{n(t)}
      \langle \chi_k \chi_l \chi_p \chi_q \rangle_{\chi_i} 
      \left\langle \frac{1-\exp(i\omega_1\tau_k)}{\omega_1} \frac{1-\exp(-i\omega_1\tau_l)}{\omega_1} 
              \frac{1-\exp(i\omega_2\tau_p)}{\omega_2} \frac{1-\exp(-i\omega_2\tau_q)}{\omega_2} \right\rangle_{\tau_i} \\
&\qquad \qquad \qquad \qquad  \langle \exp( i \omega_1 (T_k - T_l) + i \omega_2 (T_p - T_q)) \rangle_{T_i} \biggr\rangle_{n(t)}.
\end{split}
\end{equation}
This time, the non vanishing terms in the sum must have the property
$\omega_1 (T_k - T_l) + \omega_2 (T_p - T_q) = 0$ which is possible
only for $k=l$ and $p=q$:
\begin{equation}
\langle \exp( i \omega_1 (T_k - T_l) + i \omega_2 (T_p - T_q)) \rangle_{T_i} \simeq \delta_{kl} \delta_{pq}.
\end{equation}
We get with this approximation
\begin{equation}
\begin{split}
\left\langle S_t(\omega_1) S_t(\omega_2) \right\rangle
&\simeq \frac{1}{t^2} \left\langle \sum_{\substack{k,p=1 \\ k \neq p}}^{n(t)} \langle d_k(\omega_1) d_k(-\omega_1) \rangle \langle d_p(\omega_2) d_p(-\omega_2) \rangle \right\rangle
       + \frac{1}{t^2} \left\langle \sum_{k=1}^{n(t)} \langle d_k(\omega_1) d_k(-\omega_1) d_k(\omega_2) d_k(-\omega_2) \rangle  \right\rangle \\
&\simeq \frac{1}{t^2} \langle n(t) (n(t) - 1) \rangle \langle d_1(\omega_1) d_1(-\omega_1) \rangle \rangle \langle d_1(\omega_2) d_1(-\omega_2) \rangle
       + \frac{1}{t^2} \langle n(t) \rangle \langle d_1(\omega_1) d_1(-\omega_1) d_1(\omega_2) d_1(-\omega_2) \rangle \\
&\simeq \frac{1}{t^2} \langle n(t)^2 \rangle  \rangle \langle d_1(\omega_1) d_1(-\omega_1) \rangle \rangle \langle d_1(\omega_2) d_1(-\omega_2) \rangle \\
&\simeq \langle Y_\alpha^2 \rangle \langle S_t(\omega_1) \rangle \langle S_t(\omega_2) \rangle.
\end{split}
\end{equation}
This shows Eq.~\eqref{eq:condition2} and therefore with Eq.~\eqref{eq:condition1}
the equality of the second moments of Eq.~\eqref{fractionalPeriodogram}.
Similar calculations can be performed in general for higher order moments
giving rise to Eq.~\eqref{fractionalPeriodogram}.